# Emergence Of Ferromagnetism In Nanoparticles Of Antiferromagnetic $Nd_{0.4}Sr_{0.6}MnO_3$

S. Kundu and T. K. Nath

*Department of Physics and meteorology, Indian Institute of Technology,*
*Kharagpur, West Bengal, 721302, India*
*souravphy@gmail.com*

**Abstract.** We have investigated the magnetic state of $Nd_{0.4}Sr_{0.6}MnO_3$ with the variation of its grain size down to average nanometric diameter (40 nm). The bulk sample is antiferromagnetic (AFM) in nature. However, on reduction of the grain size, emergence of ferromagnetic behavior is experimentally observed. Linear and nonlinear ac magnetic susceptibility measurements reveal that ferromagnetic nature is enhanced as the grain size is reduced. Large coercivity and bifurcation between zero field cooled and field cooled magnetization curves indicate an intrinsic disorder and large anisotropy in the system. Observed behaviors are attributed to surface disorder as well as to possible pressure effect on nano grains.

## INTRODUCTION

It is well known that manganites are a particular class of materials which show a strong coupling between their lattice, spin, charge and orbital degrees of freedom [1]. Due to this strong coupling, manganites display some unusual physical behavior. It is quite expected and in many cases experimentally verified, that due to the reduction of grain size manganites show some interesting and different properties compared to their bulk counterparts [2]. We have synthesized a heavily hole doped manganite $Nd_{0.4}Sr_{0.6}MnO_3$ and performed a systematic study of its magnetic properties with the variation of grain size. The bulk $Nd_{0.4}Sr_{0.6}MnO_3$ has A - type AFM ground state [3]. The ferromagnetic layers are antiferromagnetically stacked in such fascinating type of AFM ordering (A-type). This compound is supposed to have metallic character in the ground state due to the occurrence of in plane double exchange interaction. We have shown that the nanoparticles of $Nd_{0.4}Sr_{0.6}MnO_3$ show ferromagnetic (FM) character below certain temperature. Previously, some work has been carried out on AFM materials in the nanometric form and emergence of ferromagnetism was observed due to the enhanced spin disorder in the surface of the grains [4,5]. To the best of our knowledge a study of reduction of grain size down to the nanometric scale in this manganite of $Nd_{0.4}Sr_{0.6}MnO_3$ is new in literature.

## EXPERIMENTAL DETAILS

The polycrystalline sample of $Nd_{0.4}Sr_{0.6}MnO_3$ was synthesized through chemical pyrophoric reaction route. The as prepared powder was ground thoroughly and divided into three parts and sintered at different temperatures, namely 1150 °C, 850 °C and 750 °C for 5 hrs, in air. On sintering, three polycrystalline samples of different grain size were produced. The samples are designated as NSMO1150, NSMO850 and NSMO750 (numbers indicate the sintering temperature) throughout this paper.

The characterization of the samples has been carried out through x- ray diffraction, FESEM, high resolution TEM and EDAX for compositional analysis. The detailed high precision complex ac magnetic susceptibility measurements were carried out in a homemade ac susceptometer. The signal was detected by a lock - in – amplifier (SR 830). We have also measured the higher harmonic response of ac susceptibility (nonlinear susceptibility) by detecting the signal with frequency integral multiple of the frequency of the exciting ac magnetic field (h). The dc magnetic measurements were also carried out by employing a homemade vibrating sample magnetometer. In this case also, the signal was detected by a same kind of lock - in - amplifier. During all the measurements, temperature was monitored by a precision PID temperature controller (Lakeshore, model- 325).

## RESULTS AND DISCUSSION

From the XRD patterns in Fig. 1 (a), it is evident that the samples are of single phase in nature with perovskite structure without out any impurity phase within the resolution of the diffractometer. Interestingly, one can observe that all the peaks of NSMO1150 sample are splitted into two or more. This indicates that there could be a change in crystallographic structure that has taken place due to the increase of sintering temperature from 750 °C to 1150 °C. It is also possible that the peaks for the nanometric sample have broadened to smear out the splitting. From the TEM image (Fig. 1 (b)) the calculated average grain size of NSMO750 sample is about 40 nm. The FESEM pictures show that the average grain size of NSMO850 sample is about 80 nm (Fig. 1(c), while that of NSMO1150 is more than 100 nm (Fig. (d)). The TEM and FESEM images taken at different portions of the samples show a unique nature of grain indicating the absence of impurity materials. However, the grains in all cases have a large distribution and are agglomerated. We have taken NSMO1150 sample as the bulk reference one. From the EDAX spectrum (Fig. 2 (a)), the calculated formula of NSMO750 is $Nd_{0.40}Sr_{0.63}Mn_{0.98}O_{2.98}$. This implies the authenticity of the sample composition even in the nanometric regime. Change in grain size does not affect the sample composition. The high resolution lattice image (Fig. 2 (b)) and the Selected Area Diffraction (SAD) in Fig. 2 (c) of NSMO750 show good crystallinity and polycrystalline nature of the sample. The d - spacing calculated from different images are nearly similar to that of our sample. This also supports the phase purity of the samples.

The measured real part of linear ac susceptibility ($\chi_1^R$) at an ac field of 3 Oe and 555.3 Hz frequency, of all the samples, are shown in Fig. 3 (a). For NSMO1150 sample, the $\chi_1^R$ vs. T curve show an AFM nature with a peak at around 225 K (Neel temperature). From 225 K the magnitude of susceptibility rapidly falls as the temperature is lowered further. This indicates that the AFM ordering takes place below this temperature. However, for NSMO850 and more pronouncedly for NSMO750, the magnitude of susceptibility does not fall so rapidly as the temperature is decreased. This is a clear indication of the fact that the AFM ordering has been destabilized due to the reduction of grain size. Moreover, the $\chi_1^R$ vs. T curve for the lowest particle size sample (NSMO750) shows a FM to PM like transition at around 250 K as obtained from the minimum position of temperature derivative of $\chi_1^R$ with respect to temperature (Inset of Fig. 3 (a)).

To further elucidate our claim of observing ferromagnetism due to size reduction of $Nd_{0.4}Sr_{0.6}MnO_3$, we have performed nonlinear ac susceptibility measurements on this sample. The magnetization (M) of a sample can be expanded in terms of applied field h as $M = M_0 + \chi_1 h + \chi_2 h^2 + \chi_3 h^3 + \ldots$ ($M_0$ is spontaneous magnetization) [6]. $\chi_2$, $\chi_3$ etc. are the non linear susceptibilities. The inversion symmetry of magnetization with respect to h, ($M(h) = -M(-h)$), claims that $\chi_{2,}$ $\chi_4$ etc. will have finite value if $M_0$ or ferromagnetism is present in the sample [7]. The third harmonic $\chi_3$ of a ferromagnetic sample show a positive and negative peak, below and above FM-PM transition, respectively [8]. Enhancement of the peak around 225 K in $\chi_2^R$ with the decrease of grain size and the observation of a positive to negative crossover of $\chi_3^R$ at around the same temperature, reconfirms the emergence of ferromagnetic phase in the nanoparticles of $Nd_{0.4}Sr_{0.6}MnO_3$ (Fig. 3 (b)).

The dc magnetic measurements have been carried out in terms of measuring magnetization as a function of temperature and magnetic field. Figure 4 shows a large bifurcation of the zero - field - cooled (ZFC) and field - cooled (FC) magnetization of NSMO750 sample. This kind of features like strong irreversibility between ZFC - FC magnetization is generally observed in a superparamagnet or in spin/cluster glass like materials. The nature of ZFC - FC magnetization of NSMO750 sample reveals that the observed ferromagnetism in it cannot be labeled as a conventional one. A high degree of frustration and disorder is present in the sample. The magnetic hysteresis curves (M vs. H) of the NSMO750 sample, recorded at different temperatures, are shown in the inset of Fig. 4. The M-H behavior displays a ferromagnetic nature with large coercive field (≈100 Oe) in the low temperature region. The nature of FC curve shows a saturating nature with the decrease of temperature and shows no tendency of going up in magnitude in the low temperature region. This indicates that the system NSMO750 is an interacting system of particles. This is quite likely to be observed, as the grains or particles are overlapping or very closely packed in nature in our samples. However, we believe that this overlapping or connected nature of grains does not lessen general nanometric effects as the grain size itself has been reduced by an order of magnitude compared to the bulk counterpart. Some effects may arise in the electronic transport properties of the system.

To explain all the observed experimental results, we first consider the effect of enhanced grain boundary in the nanoparticles. The surface region of a nano grain consists of lot of defects and broken bonds as the crystallographic order terminates there. The effect of reducing the grain size in an AFM material is the enhanced disorder in the grain boundary region which will weaken the AFM exchange interaction giving rise to a net moment of a grain due to the uncompensated spins at the surface [4]. It seems that the disordered surface of our nanometric samples is an obvious origin of the observed ferromagnetic behavior. However, due to the presence of large number of defects in the grain boundary region, the spins will experience a fair amount of random anisotropic forces due to the pinning at those defects. The large bifurcation between the ZFC and FC magnetization and large coercivity are the consequences of this anisotropy forces competing with the thermal energy [9].

Furthermore, we want to mention another possibility of emergence of ferromagnetic behavior on reduction of the grain size. This is precisely the structural change introduced by the high surface pressure in the nano grains. Recently, it has been shown that the pressure on a grain due to the surface tension is quite high in the nanometric regime [10]. This high pressure may produce some intrinsic structural change in the bulk of the grains. From the XRD pattern we have already observed that a possible modification in the crystallographic structure has taken place in the nanometric NSMO750 sample. It has already been shown that the external hydrostatic pressure on a manganite can change the existing ground state and modify the physical properties [11]. In our case, a small structural change can produce a ferromagnetic behavior as the ferromagnetic order already existing in the individual planes. This will also produce a phase separation (AFM and FM) in the bulk of the nanomaterials. Hence, it appears that, beside the surface effect the intrinsic change in the magnetic state may be responsible for the observed ferromagnetic behavior in the nanoparticles of $Nd_{0.4}Sr_{0.6}MnO_3$. However, it is to be mentioned here, that a detailed quantitative structural characterization has to be carried out to confirm such an effect. Hence, on the basis of our present study, we stress upon the surface effects which is very common to nanomaterials, to describe the observed nanosize effects.

## CONCLUSIONS

A detailed investigation of the magnetic properties through ac susceptibility and dc magnetic measurements has been carried out on samples of $Nd_{0.4}Sr_{0.6}MnO_3$ with different grain size down to nanometer scale, synthesized by chemical route. The bulk sample is A-type antiferromagnet in the ground state. As the grain size is reduced an enhanced ferromagnetic behavior is observed. Nonlinear susceptibility provides strong evidences of the ferromagnetic nature by showing the existence of spontaneous magnetization in the nanoparticles. We have attributed the observed results to surface disorder in the nanograins and a possible increase in surface pressure which is likely to produce a structural change and phase separation in this AFM manganite $Nd_{0.4}Sr_{0.6}MnO_3$.

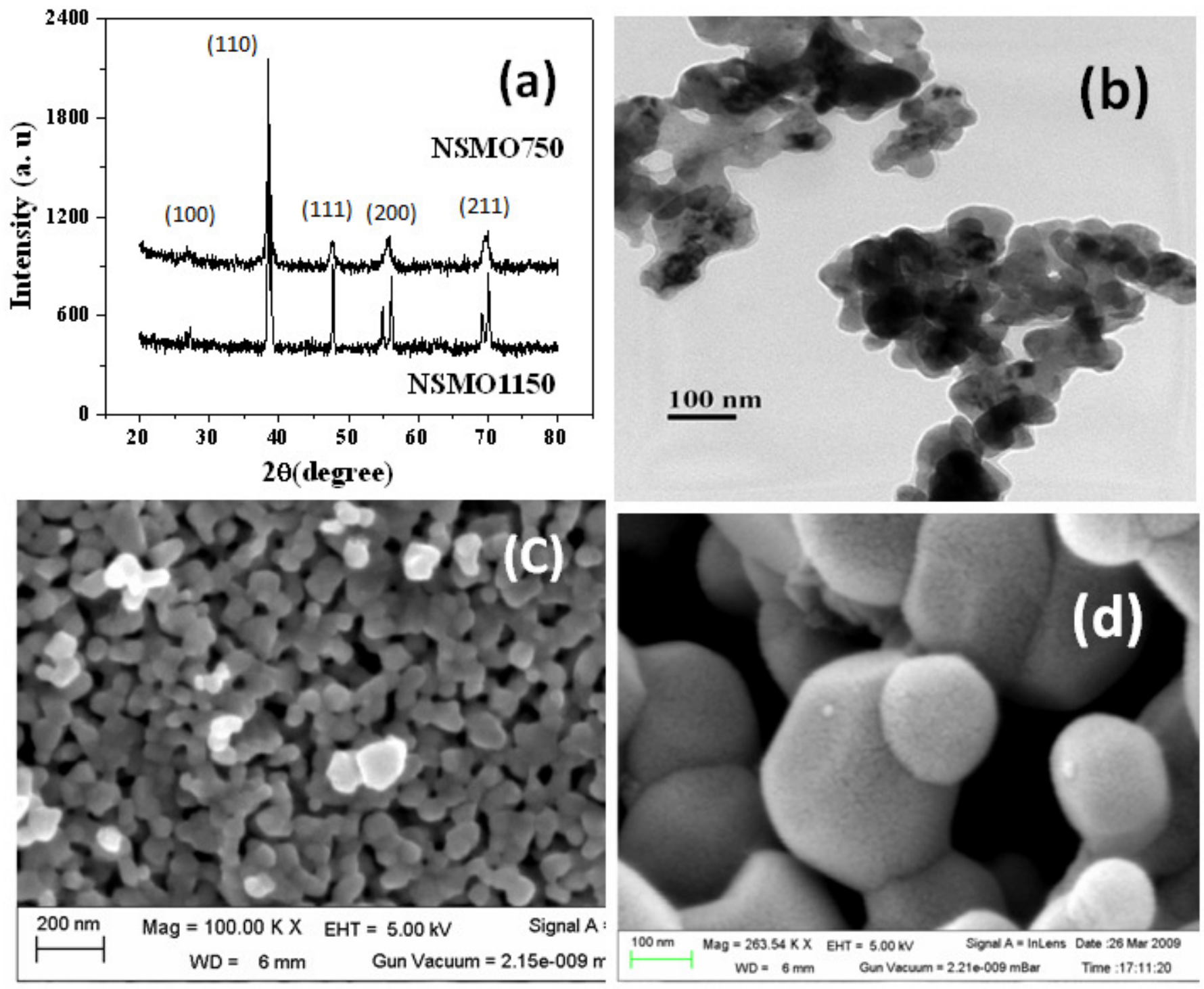

**FIGURE 1.** (a) XRD micrographs of the samples. (b) TEM image of NSMO750 sample. (c) FESEM image of NSMO850 sample. (d) FESEM image of NSMO1150 sample.

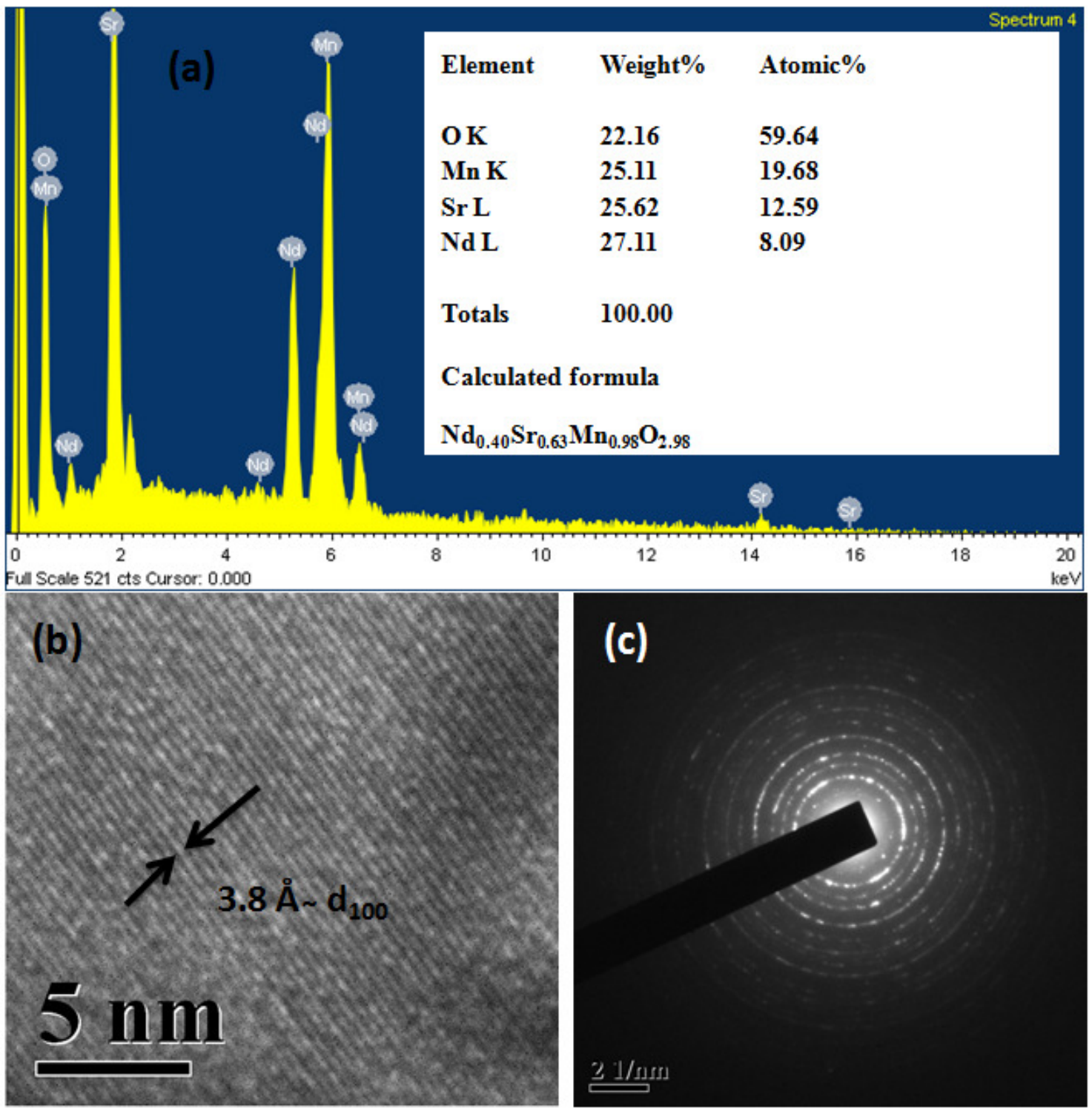

**FIGURE 2.** (a) The EDAX spectra of NSMO750. (b) High resolution lattice image. (c) SAD pattern.

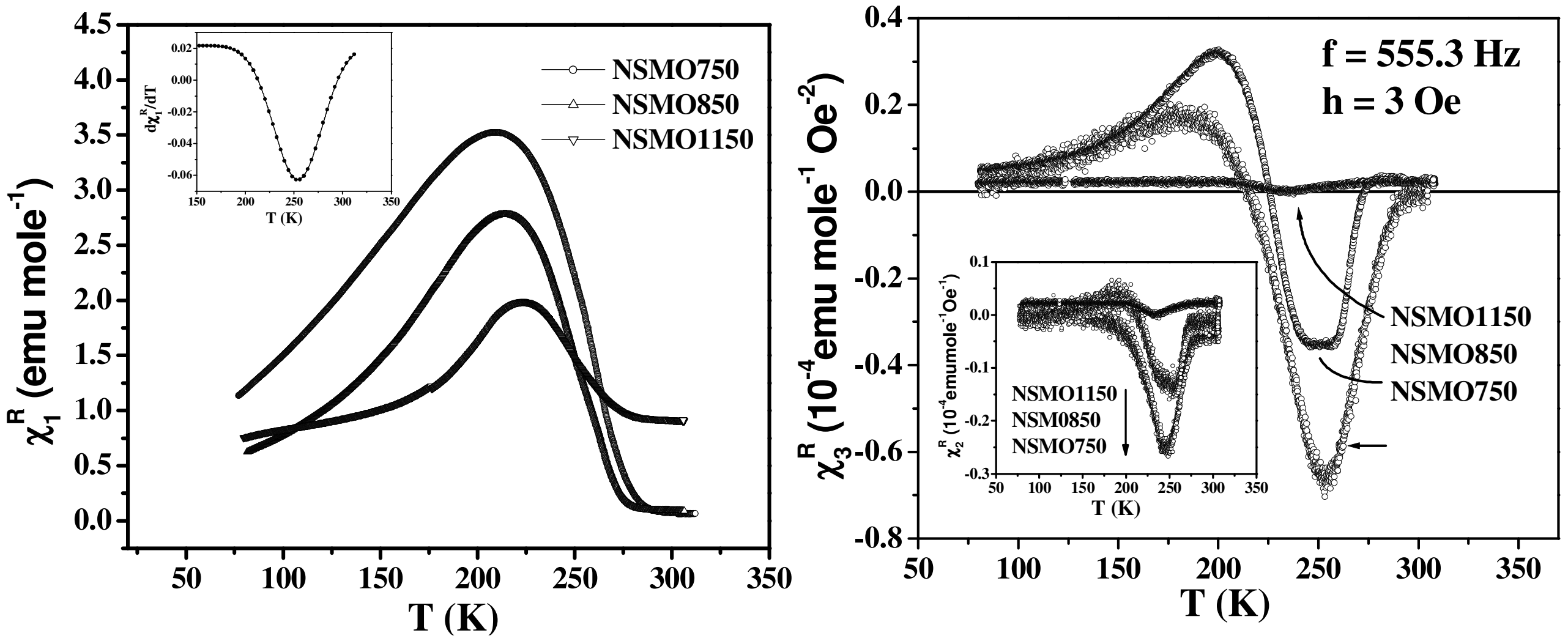

(a) (b)

**FIGURE 3.** (a) Temperature variation of linear susceptibility ($\chi_1^R$) of the samples. Inset shows the temperature derivative of $\chi_1^R$ plotted as function of temperature of NSMO750 sample. (b) Variation of the third order nonlinear susceptibility ($\chi_3^R$ ) with temperature of all the samples. Inset shows the measured second order susceptibility ($\chi_2^R$ ) of all the samples.

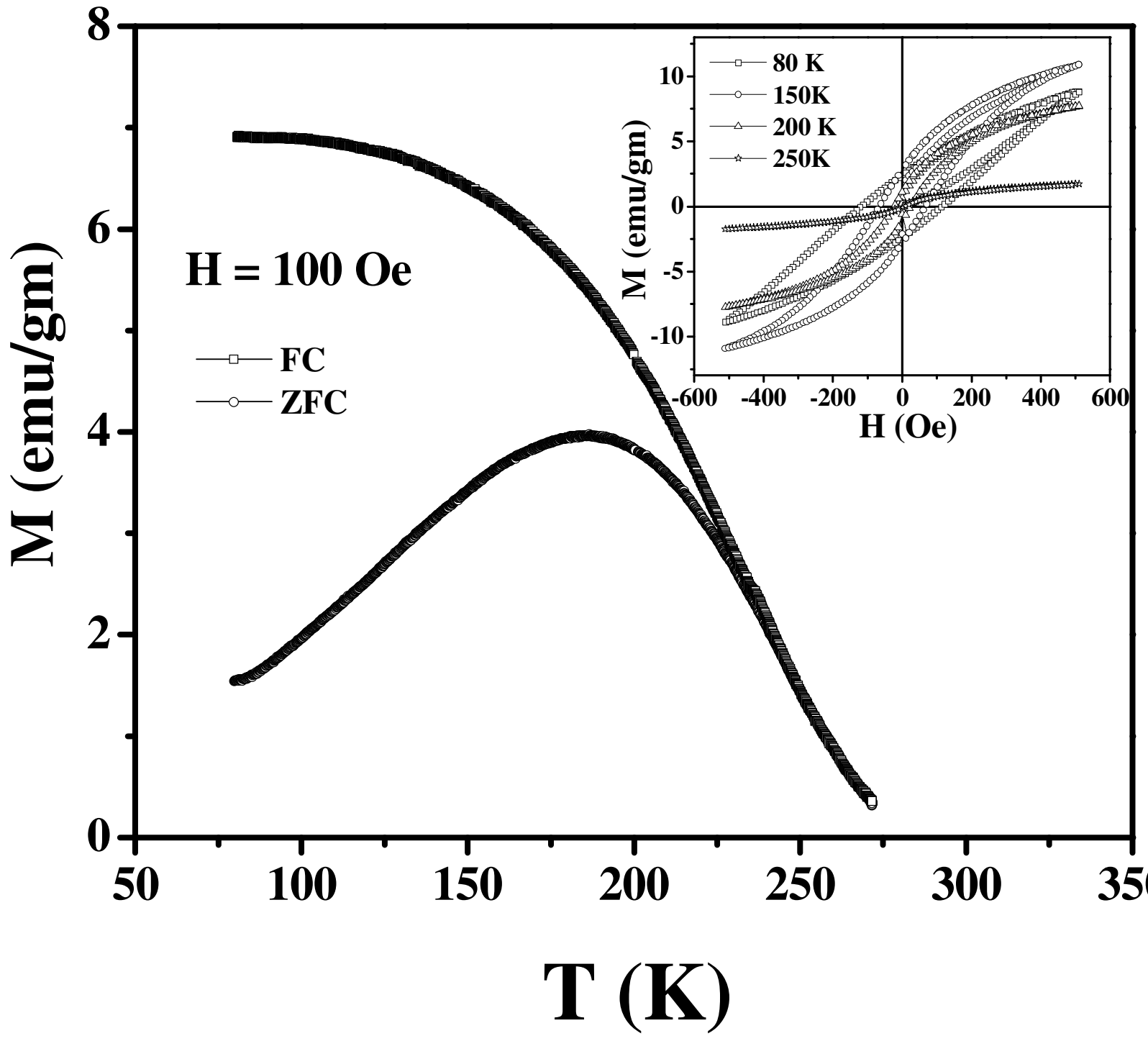

**FIGURE 4.** Field cooled and zero field cooled magnetization of NSMO750 sample measured at 100 Oe. Inset shows the M-H curves of the sample at different temperatures.